\documentclass[structabstract]{aa}  
\usepackage{color}
\usepackage{lscape}
\usepackage{graphicx}
\usepackage{natbib}
\begin{document} 

\title{The scaling of X--ray variability with luminosity in Ultra-luminous
X-ray sources}

   \author{O. Gonz\'alez-Mart\'in\inst{1,2}\fnmsep\thanks{\email{omaira@physics.uoc.gr}}
	  I. Papadakis\inst{2,1}
	  P. Reig\inst{1,2}
          \and  A. Zezas\inst{2,1}
          }

   \institute{$\rm{^1}$ IESL, Foundation for Research and Technology, 711 10, Heraklion,
	      Crete, Greece\\
	      $\rm{^2}$ Physics Department, University of Crete, P.O. Box 2208, Gr-710 03
   Heraklion, Crete, Greece\\
}

   \date{Received March 30th, 2010}
\authorrunning{O. Gonz\'alez-Mart\'in et al.}
\titlerunning{X-ray variability versus luminosity of ULXs}
  \abstract
   {}
   {To investigate the relationship between the X-ray variability  amplitude and
   X-ray luminosity for a  sample of 14 bright Ultra-luminous X-ray sources (ULXs) with \emph{XMM-Newton}/EPIC
   data, and compare it with the well established similar relationship for Active Galactic Nuclei (AGN).}
   {We computed the normalised excess variance in the 2--10 keV light  curves of
   these objects and their 2-10 keV band  intrinsic luminosity
   $\rm{L_{2-10~keV}}$. We also determined model  ``variability-luminosity"
   relationships for AGN, under several assumptions regarding their 
   power-spectral shape. We compared these model predictions at low 
   luminosities with the ULX data.}
   {The variability amplitude of the ULXs is significantly smaller than that expected 
   from a simple extrapolation of the AGN ``variability-luminosity" relationship at  
   low luminosities. We also find evidence for an anti-correlation between the  
   variability amplitude and $\rm{L_{2-10~keV}}$ for ULXs. The shape of this  
   relationship is consistent with the AGN data but only if the ULXs data are shifted 
   by four orders of magnitudes in luminosity.  }
   {Most (but not all) of the ULXs could be `scaled-down' version of AGN if we 
   assume that: i) their black hole mass and accretion rate are of the order of 
   $\rm{\sim(2.5-30)\times 10^3M_{\odot}}$ and $\rm{\sim 1-80}$\% of
   the Eddington limit, and  ii) their Power  Spectral Density has a doubly
   broken power-law shape. This PDS shape and  accretion  rate is consistent
   with Galactic black hole systems operating in their  so-called  ``low-hard"
   and ``very-high" states.}
   \keywords{Black hole physics -- X-ray: galaxies -- X-ray: binaries
               }

   \maketitle
%
 
\section{Introduction}

Ultra-luminous X-ray sources (ULXs) are point-like sources with luminosities 
greater than $\rm{10^{39}~erg~s^{-1}}$ in the 0.3-10 keV band.  This high 
luminosity is greater than that expected from stellar mass black holes (M$_{\rm
BH} \rm{< 20~M_{\odot}}$) accreting at the Eddington limit. Because they are
usually located away from the nucleus of the galaxies they are unlikely to   be
associated with super massive black holes (SMBH, M$_{\rm BH} \rm{>
10^{5}~M_{\odot}}$), typically observed at the centre of  active galactic nuclei
(AGN). Their high luminosities can be explained if we assume that they host a
black hole (BH) with an ``intermediate mass",  around  100-10000
$\rm{M_{\odot}}$ \citep[IMBHs,][]{Colbert99}. However, the true nature of these
sources is still unclear and  other mechanisms as anisotropic emission
\citep{King01} or accretion onto the BH in excess of the expected Eddington
limit \citep{Begelman02} could also explain them \citep[see ] [for a recent
review]{Roberts07}. The question of what powers ULXs will be conclusively
answered by a direct mass measurement based on the determination of the binary
orbit. However, due to their extragalactic nature, the study of the ULX
counterparts in other bands has been difficult. 

In the meantime, both spectral and timing methods have been used over the last
few years in an attempt to constrain the mass of the compact object in ULXs 
\citep{Miller04}. A common spectral method uses the temperature and luminosity
of the accretion disc emission to determine the BH mass (assuming the standard
Shakura-Sunyaev models). As for the timing methods, one can either use the
\citet{McHardy06} and \citet{Kording07} scaling relationships of the
characteristic time scales in Galactic black hole binaries (GBHs) and SMBHs with
BH mass and accretion rate, or the  timing -- spectral scaling for the QPOs in
GBHs \citep[e.g.][]{Shaposhnikov07, Shaposhnikov09}. The results from the
application of this method to ULXs have been  non-conclusive. Some studies
suggest that ULXs are stellar mass BHs \citep[e.g.][]
{Gladstone09,Roberts07,Zezas07,Dewangan06} while others imply that they host 
IMBHs \citep[e.g.][]{Casella08,Miller04}. 

In this paper we investigate the relationship between the X-ray variability 
amplitude and luminosity for a sample of 14 bright ULX sources using
\emph{XMM-Newton}/EPIC data. Our first aim is to measure the 2--10 keV
normalised  excess variance, $\rm{\sigma^{2}_{NXS}}$, for these objects and
investigate whether it correlates with the source luminosity. The normalised
excess variance is a simple-to-calculate quantity that measures the intrinsic
variability amplitude of a source. It can be a useful complementary tool to the
full-blown power-spectrum density (PSD) analysis, and it has the advantage that
it can be applied to a larger number of objects as it does not require
high-quality data (i.e. long, high signal-to-noise light curves).  

%
\begin{table*}
\caption{The sample and observational details.}             
\label{tab:observations}      
\centering          
\begin{tabular}{l c c c c c c c c c}     
\hline\hline       
           &          &            &    &          &     &  &       &   & \\
Name & R.A.    & Dec     & Dist. & ObsID &  Mode$^*$ & Filter & Back. Radius &Seg.($T_{\rm net}$) \\  
     & (J2000) & (J2000) & (Mpc) &	 &       &        & (pixels)    & 	(ksec)  \\ 
\hline                     
           &          &            &    &          &     &  &       &   & \\
NGC55ULX   &00 15 28.9&-39 13  19.1&1.9 &028740201& FW& Thin1  &534&1(30)\\
NGC253PSX-2&00 47 32.9&-25 17  50.3&3.2 &152020101& FW& Thin1  &500&2(40,35)\\
           &          &            &    &125960101& FW& Medium &500&1(39)\\
NGC1313X-1 &03 18 20.0&-66 29  11.0&4.0 &106860101& FW& Medium &930&1(31)\\
           &          &            &    &405090101& FW& Medium &677&3(40,38)\\
NGC1313X-2 &03 18 22.3&-66 36  03.8&4.0 &106860101& FW& Medium &500&1(31)\\
           &          &            &    &405090101& FW& Medium &500&2(40,38)\\
NGC2403X-1 &07 36 25.6&+65 35  40.0&3.5 &164560901& FW& Medium &556&2(40,30)\\
HoIIX-1    &08 19 29.0&+70 42  19.3&3.3 &200470101& FW& Medium &998&1(37)\\
M81X-6     &09 55 32.9&+69 00  33.3&3.7 &111800101& SW& Medium &500&2(40,37)\\
M82X-1     &09 55 50.2&+69 40  47.0&4.0 &206080101& FW& Medium &500&2(40,37)\\
HoIXX-1    &09 57 53.2&+69 03  48.3&3.3 &200980101& LW& Thin1  &737&2(40,37)\\
NGC3628X-1 &11 20 15.8&+13 35  13.6&12.0&110980101& EFW &Thin1 &500&1(38)\\
NGC4559X-1 &12 35 51.7&+27 56  04.1&8.9 &152170501& FW& Medium &1038&1(37)\\
NGC4945X-2 &13 05 33.3&-49 27  36.3&4.0 &204870101& FW& Medium &500&1(40)\\
NGC5204X-1 &13 29 38.6&+58 25  05.7&5.3 &405690201& FW& Medium &831&1(37)\\
NGC5408X-1 &14 03 19.6&-41 22  59.6&4.9 &500750101& FW& Thin1  &500&1(40)\\
           &          &            &    &302900101& FW& Thin1  &500&2(40,40)\\
           &          &            &    &	  &   &        &   & \\ \hline
           &          &            &    &	  &   &        &   & \\
POX52      &12 02 56.9&-20 56  03.3&96.1&302420101& FW& Medium &500&2(40,40)\\
           &          &            &    &         &   &        &   & \\
\hline                  
\end{tabular}\\
$~$\newline
\scriptsize{$^*$ 'FW', `SW', `LW" and 'EFW' stand
for ``Full", ``Small", ``Large" and ``Extended Full Window" mode of the PN
detector, respectively.}
\end{table*}

It is well established that  $\rm{\sigma^{2}_{NXS}}$ is anti-correlated with
luminosity in AGN \citep{Nandra97,Leighly99,Turner99}. Moreover, the excess
variance  anti-correlates with BH mass \citep{Lu01,Bian03,O'Neill05,Miniutti09,
Zhou10}. Furthermore, \citet{Papadakis04} showed that the ``variability--mass" 
relationship is probably the physically fundamental relationship  rather than
the ``variability-luminosity" relationship in these objects. Our second aim is
to compare the ``variability--luminosity" relationship for ULXs with that of AGN
with known BH mass. Following \citet{Papadakis04}, if we assume a
\emph{universal} PSD shape for AGN, which scales appropriately with BH mass and
accretion rate, we can then make predictions on the expected AGN
``variability--luminosity" relationship at low luminosities. We want to
investigate whether the ULX ``variability--luminosity" data are consistent with
various model ``variability--luminosity" relationships for AGN, and if yes, what
are the implications for the ULX PSD shape, BH mass and accretion rate. 

In Section 2 we present the sample selection. In  Section 3 we describe the data
reduction, and in  Section 4 we discuss the data analysis and present our 
results. We present a short discussion of their implications and  our
conclusions  in Sections 5 and 6, respectively.


\section{The sample}\label{sec:sample}

We considered all bright ULXs reported in the literature, and in particular the
objects studied by \citet{Heil09} and \citet{Gladstone09}. The \citet{Heil09}
sample includes all bright ULXs which have been observed with \emph{XMM-Newton}
for more than 25 ksec and their 0.2--10 keV flux is greater than $\rm{5\times
10^{-13}erg~cm^{-2}s^{-1}}$. The \citet{Gladstone09} sample includes all ULXs
observed  with EPIC/\emph{XMM-Newton} with more than 10000 net counts in the 
0.3--12 keV EPIC band. 

One of our main aims is to compare the 2--10 keV variability amplitude of ULXs
with the variability amplitude of the nearby AGN studied mainly by
\citet{O'Neill05}.  The length, \emph{T}, and the bin size, $\Delta t$, of a
light curve  determines the lower and higher frequency sampled, since $\nu_{\rm
min}=1/T$ and $\nu_{\rm max}=1/(2 \Delta t)$ Hz.  The excess variance of the
light curve depends on the intrinsic power-spectrum and also on the minimum and
maximum frequencies (see Section 4.5). For that reason  we used the same length
for the light curves as those of  \citet{O'Neill05}. Thus, we considered light
curve segments with a length of $30-40$ ksec. Regarding $\Delta t$, due to the
low count rate of all objects in the sample,  we used bins of size 1000 s in
order to increase the signal-to-noise of their light curves.

Consequently, we chose from the \citet{Heil09} and \citet{Gladstone09} samples
those sources which  were observed by \emph{XMM-Newton} with a net exposure,
$T_{\rm net}$,  larger than 30 ksec. For this reason we did not consider  the
\emph{XMM-Newton} data of  M\,33 X--8, IC\,342 X--1,  NGC\,4395 X--1, and M\,83 ULX.
We did not consider NGC\,4395 X--1 either because it was located on a gap of the
PN detector during its \emph{XMM-Newton} observation with  $T_{\rm net}>30$
ksec.

Our final sample comprises of 14 ULXs. Table \ref{tab:observations} lists their
coordinates, distance, and the \emph{XMM-Newton} observation details. 
Coordinates and distances were taken from the 
NASA/NED\footnote{http://nedwww.ipac.caltech.edu} database. Distances 
correspond to the average redshift-independent estimate for each object. For
four sources (namely NGC\,253 PSX-2, NGC\,1313 X--1, NGC\,1313 X--2, and NGC\,5408
X--1) we were able to retrieve from the archive two observations with $T_{\rm
net}$ larger than 30 ksec. Note that the starburst galaxy M\,82 contains two
ULXs, namely X41.4+60 and X42.3+59, which are unresolved by  \emph{XMM-Newton},
and they both contribute to the M\,82 X--1 light curve. During the 2004 April
observation that we considered in this work, approximately 84\% of the observed
count rate originates from X41.4+60 \citep{Feng07}.

In order to extend the \citet{O'Neill05} sample to include AGN with low BH
masses, we also  considered the \emph{XMM-Newton} observation of POX\,52, which
hosts an AGN with a low BH mass \citep[$\rm{M_{BH}}=1.6\times10^{5}~M_{\odot}$; 
][]{Barth04}.  The coordinates, distance, as well as the  \emph{XMM-Newton}
observation details for this source are also listed in
Table~\ref{tab:observations} (the distance in this case corresponds to the
``luminosity distance" estimate of NASA/NED).

\section{Data reduction}\label{sec:reduction}

Data were retrieved  from the \emph{XMM-Newton} public data 
archive\footnote{http://xmm.esac.esa.int/xsa/index.shtml}. We used the 
\emph{XMM-Newton} Science Analysis System
SAS\footnote{http://xmm.esac.esa.int/sas/} software version 9.0.0 and followed
standard procedures to extract  science  products from the Observation Data
Files (ODFs). 

We used data from  the EPIC-pn camera only due to its superior statistical
quality. Source counts in each case were  accumulated from a circular region of
radius 400 pixels, centred on the source's  RA and Dec. In the case of NGC\,253
PSX-2 (ObsID 152020101) we used a radius of 300 pixels to avoid contamination
from a nearby source and the detector gap. Background data  were extracted from
a source free circular region on the same CCD chip than the source  (background
region radii are listed in Column 8 of Table~\ref{tab:observations}). We
selected only single and double pixel events (i.e. patterns of 0-4). Bad pixels
and events too close to the edges of the CCD chips were rejected using
``FLAG=0." Given the observed count rate, photon pile-up is negligible for the
PN detector in all cases.

Source and background light curves in the 2--10 keV band  were extracted using
{\sc evselect} task on SAS with a 1000-sec bin. They were screened for high
background (usually at the end and/or the beginning of the individual
observations)  and flaring activity. After rejection of the respective time
intervals, the total useful observation time for each observation is usually
less than the  original PN exposure time. We chose to study only those light
curves with at least one ``clean" segment longer than 30 ksec.

\begin{figure}[!t]
\centering
\includegraphics[width=1.0\columnwidth]{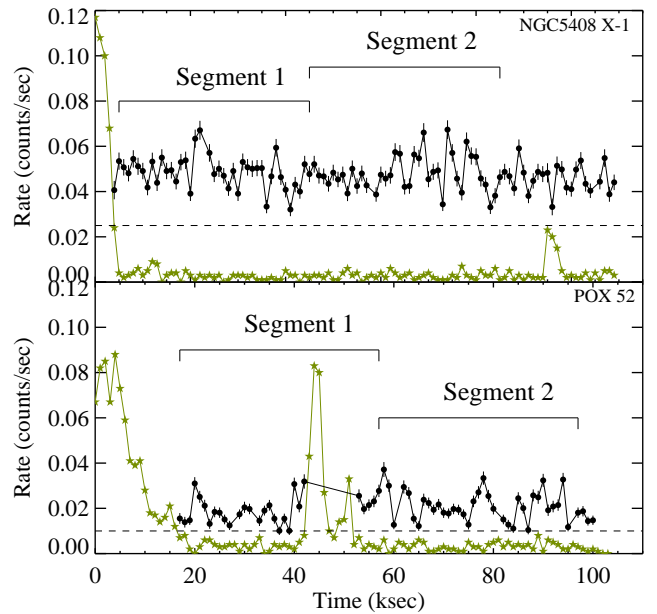}

\caption{Light curves of NGC\,5408 X1/ObsID 302900101 (top panel) and POX\,52
(bottom panel). Stars indicate the background light curves and filled dots 
indicate the background subtracted light curves, plotted only for  the
observation period for which the background activity is ``low" (see text for
details). The brackets on top of the light curves indicate the  segments that
were used to estimate the excess variance in each case.}

\label{fig:lightcurves}
\end{figure}

As an example, in the top panel of  Fig.~\ref{fig:lightcurves} we show the light
curve of  NGC\,5408 X--1, which is typical of the light curves of all sources in
the sample. Filled circles and stars indicate the  2--10 keV (background
subtracted) source and background light curves, respectively.  The background
light curve corresponds to the full PN exposure length while the source light
curve is plotted only for those parts of the observation when the background
activity  was ``low''. In general, as background ``loud'' we identified the
observation  parts where:  i) the background light curve showed ``flare"-like
events  and/or prominent decreasing/increasing trends (usually  at the start/end
of an observation), and ii)  the ``net" source count  rate was less than twice
the background count rate. The brackets on top of the  NGC\,5408 X--1 light
curves indicate the light curve segments we chose for this observation. Clearly,
the background light curve is stable, and of much less intensity than the ``net"
light curve count rate. This was the case for almost all of the light curve 
segments we used in this work. We also extracted the EPIC-pn spectra, after the
rejection of the time intervals affected by high background,  using  single and
double  events (PATTERN $\rm{<}$=4). Response and auxiliary matrices were
created with SAS tools {\sc rmfgen}  and {\sc arfgen}, respectively.

The POX\,52 \emph{XMM-Newton} data were reduced in the same way. Its 2--10 keV
(background subtracted) source and background light curves  are also plotted in
Fig.~\ref{fig:lightcurves} (bottom panel). The observation is affected by high
background flaring activity during the  first $\sim 20$~ksec, and after $\sim
40$ ksec since the start of the observation, which lasted for almost 10~ksec
(note that this observation shows the worse background ``flaring" activity among
all the light curves we studied in this work). 

\section{Data analysis and results}\label{sec:analysis}

\subsection{The variability amplitude estimation}

As a measure of the intrinsic variability amplitude of the light curves  we
computed their normalised excess variance, $\rm{\sigma^{2}_{NXS}}$. Its square
root is a measure of  the average variability amplitude of a source as a
fraction of the light curve mean. We used the prescription given by
\citet{Vaughan03} to estimate $\rm{\sigma^{2}_{NXS}}$  and its error,
$\rm{err(\sigma^{2}_{NXS})}$\footnote{The error accounts only for the
uncertainty due to the Poisson noise and not due to the stochastic character of
the intrinsic variability process.}, as follows:

\begin{equation}  
\sigma^{2}_{\rm NXS}=\frac{S^{2}-<\sigma_{err}^{2}>}{<x>^2},  
\end{equation}

\begin{equation}
{\rm err}(\sigma^{2}_{\rm NXS})=\sqrt{\frac{2} {\rm N} ( \frac
{<\sigma^{2}_{\rm err}>} {<x>^2} )^2+ \frac {<\sigma^{2}_{\rm err}>} {\rm
N} \frac{4\sigma^{2}_{\rm NXS}}{<x>^2}},
\end{equation}

\noindent where  $x$, $\sigma_{err}$, and $N$ are the count rate,
its error, and the number of points in the light curve, respectively 
and $\rm{S^{2}}$ is the variance of the light curve, i.e.:

\begin{equation}
\label{eqn:variance}
S^{2} = \frac{1}{N-1} \sum_{i=1}^{N} (x_{i} - <x>)^{2}.
\end{equation}

\noindent Following \citet{O'Neill05}, we computed $\rm{\sigma^{2}_{NXS}}$ for
each continuous light curve segment with $30\leq T_{\rm net} \leq 40$ ksec.  For
light curves longer than 40 ksec we considered only the first 40 ksec. If there
were more than one segments of duration $T_{\rm net}$ larger than  30--40 ksec
we computed $\rm{\sigma^{2}_{NXS}}$ for each one of them. The number of light
curve segments in each observation, and their $T_{\rm net}$, are listed in Table
\ref{tab:observations} (Column 9). A few ``missing" points within each segment,
due to the presence of background flaring activity, appear in one of the light
curve segments of NGC\,4559 X--1, NGC\,4945 X--2, NGC\,253 PSX-2  (ObsID
152020101), and HoII X--1. Missing points are typically  less than 10--15\% of
the total number of points. The first segment of POX 52 light curve shows the
largest number of ``missing" points (20\% of the total). The presence of missing
points in these segments should increase the uncertainty of the resulting 
$\rm{\sigma^{2}_{NXS}}$ estimates. 

Our $\rm{\sigma^{2}_{NXS}}$ estimates, for each light curve segment, are
listed in Table \ref{tab:results} (Column 3). The numbers in parenthesis in
Table \ref{tab:results} indicate the weighted  mean $\rm{\sigma^{2}_{NXS}}$
and its error in the case we had more than one excess variance estimate for the
same source. 

For two sources  the  $\rm{\sigma^{2}_{NXS}}$ measurement was negative. In
these cases, we estimated the 90\% upper limits of the intrinsic 
$\rm{\sigma^{2}_{NXS}}$ values using the 90\% upper limits on the source variance, as listed by
\citet{Vaughan03} in their Table~1. In order to constrain as much as
possible the ``variability-luminosity" ULXs correlation (see Section 4.3 below),
we assumed a PSD slope of $\rm{-}$1, and the upper limit from the \citet{Vaughan03}
simulations with the longest light curves (any other choice would result to an
even larger 90\% limit). In order to take into account the uncertainty on
$\rm{\sigma^{2}_{NXS}}$ due to the experimental Poisson fluctuations as well, we
added to these limits the value of 1.282~err($\rm{\sigma^{2}_{NXS}}$). Our final
estimates of the 90\% confidence limits for these two sources  are listed in a 
parenthesis in Table \ref{tab:results}.

Regarding M\,82 X--1, \citet{Feng07} estimate that X41.4+60 contributes  more
than 80\% of the observed count rate during the 2004 April \emph{XMM-Newton}
observation of  M\,82. X42.3+59 is highly variable source, but on time scales of
years \citep[see Fig.~5 in][]{Feng07}. On shorter time scales, the same authors
show that the PSD of X41.4+60 has a significantly higher amplitude than the PSD
of X42.3+59. 

%
\begin{table}
\caption{The excess variance ($\rm{\sigma^{2}_{NXS}}$) and 2-10 keV 
intrinsic luminosity ($\rm{L_{2-10~keV}}$) in logarithmic scale.}             
\label{tab:results}      
\centering          
\begin{tabular}{l r c c c c}     
\hline\hline       
     &       &                              &                          \\
Name & ObsID/Seg.  & $\rm{\sigma^{2}_{NXS}}$   &log(L$\rm{_X}$)      \\
     &       &       ($\times 10^{-3}$)      &                       \\  \hline
     &       &                               &           		      \\
    NGC55ULX& 028740201/1 & $\rm{147\pm 2.1}     $ & 38.7    \\
 NGC253PSX-2& 152020101/1 & $\rm{ 5.7\pm 2.4}   $ &  39.3    	 \\
            &          /2 & $\rm{-1.8\pm 2.6}   $ &            \\
	    & 125960101/1 & $\rm{ 10.0\pm 2.0}  $ &  39.4    	 \\
	    &             & $(\rm{ 6.5\pm 1.3}) $ &  (39.4)      \\	  
  NGC1313X-1& 106860101/1 & $\rm{  2.4\pm 1.6}$   &   39.6   	 \\
            & 405090101/1 & $\rm{  4.0\pm 1.5}$   &   39.6   	\\
            &          /2 & $\rm{ -0.3\pm 1.5}$   &              \\
	    &             & $(\rm{2.0\pm 0.9})$   &  (39.6)      \\
  NGC1313X-2& 106860101/1 & $\rm{ -7.1\pm 6.8}$   &  39.1    \\
            & 405090101/1 & $\rm{ 22.9\pm 1.4}$   &  39.7    \\
            &          /2 & $\rm{ 11.3\pm 1.3}$   &               \\
	    &             & ($\rm{16.2\pm 0.9}$)  &  (39.5)  	  \\
  NGC2403X-1& 164560901/1 & $\rm{ -2.3\pm 3.5}$   &  39.2    \\
            &          /2 & $\rm{ -2.8\pm 4.5}$   &               \\
	    &             & $(<8.3)$              &               \\
     HoIIX-1& 200470101/1 & $\rm{  1.3\pm 1.0}$   &  39.6    \\
      M81X-6& 111800101/1 & $\rm{  0.9\pm 2.2}$   &  39.5    \\
            &          /2 & $\rm{  8.2\pm 2.1}$   &               \\
	    &             & ($\rm{  4.9\pm 1.5}$) &               \\
      M82X-1& 206080101/1 & $\rm{  1.0\pm 0.3}$   &  40.4     \\
            &          /2 & $\rm{  0.8\pm 0.3}$   &               \\
	    &             & ($\rm{  0.9\pm 0.2}$) &               \\
     HoIXX-1& 200980101/1 & $\rm{  0.1\pm 0.6}$   &  39.8    	\\
            &          /2 & $\rm{  1.2\pm 0.6}$   &    &     	\\
	    &             & ($\rm{  0.6\pm 0.4}$) &               \\
  NGC3628X-1& 110980101/1 & $\rm{  0.6\pm 6.2}$   &  40.0    	  \\
  NGC4559X-1& 152170501/1 & $\rm{ 13.0\pm 6.6}$   &  39.6    	  \\
  NGC4945X-2& 204870101/1 & $\rm{-10.5\pm 9.5}$   &  39.0    	  \\
  	    &             & ($<  27.0$)           &               \\
  NGC5204X-1& 405690201/1 & $\rm{  1.6\pm 2.5}$   &  39.6    	  \\
  NGC5408X-1& 500750101/1 & $\rm{ 16.0\pm 5.1}$   &  39.3    	  \\
            & 302900101/1 & $\rm{  2.4\pm 5.0}$   &  39.5    	  \\
            &          /2 & $\rm{ 12.4\pm 5.1}$   &    &     	  \\
	    &             & ($\rm{10.2\pm 3.0}$)  &  (39.4)     \\
            &             & 			  &               \\ \hline
            &             & 			  &               \\
       POX52& 302420101/1 & $\rm{ 78\pm 17}$      &  40.7            \\
            &          /2 & $\rm{105\pm 15}$      &          	   \\
	    &             & $(\rm{ 93\pm 11})$      &        	  \\
\hline                  
\end{tabular}
\end{table}

\subsection{The hard band X--ray luminosity estimation}

To estimate the X-ray luminosity for each source we fitted their spectra  with
an absorbed power-law model in the 2--10 keV band. For the Galactic absorption,
we fixed the N$_{\rm H}$ values at the values derived from the HI maps of
\citep{Dickey90}. The spectral fitting was performed using XSPEC version 12.5.1.
Using the best-fit results we estimated the source flux in the 2--10 keV band,
and hence the source luminosity, $\rm{L_{2-10~keV}}$, adopting the distance
estimates listed in Table~\ref{tab:observations}. The unabsorbed
$\rm{L_{2-10~keV}}$ estimates are listed in Table~\ref{tab:results}   (Column
4)\footnote{Note that we considered a single spectrum for each observation (i.e.
we accumulated all the data for the whole ``background-quiet" period of each
observation), irrespective of the number of segments that we used for the
estimation of the normalised excess variance.}. The values in parenthesis
correspond to the mean  $\rm{L_{2-10~keV}}$ estimates, in the case there were
more than one spectrum for an object.

In half of the cases, the best-fit  $\chi^{2}_{\rm red}$ values were larger than
$\sim 1.2$. This is mainly  due to the presence of additional complexity in the
spectra that the simple power-law model cannot account for  \citep[e.g. the
presence of ``breaks" in the high energy spectra of these
sources,][]{Gladstone09}. Nevertheless, the power law model describes adequately
the broad shape of the source spectra in all cases, and the resulting  best-fit
flux measurements should be an accurate  estimate of the source X--ray {\it
continuum} flux.  To investigate this issue further, we used the 2--10 keV band best-fit 
results of \citet{Stobbart06} to estimate the 2--10 keV luminosity for the nine
sources in common. We found that
$\rm{L_{\rm ours} = L_{\rm literature}}$ in all cases except for NGC\,55 ULX, Ho\,II X--1, and Ho\,IX X--1 sources where
$\rm{L_{\rm ours}/L_{\rm literature}}=1.1$, $\rm{L_{\rm ours}/L_{\rm
literature}}=0.8$, and $\rm{L_{\rm ours}/L_{\rm literature}}=0.8$, respectively. We are thus confident that our luminosity
estimates are reliable. Regarding M\,82 X--1, \citet{Feng07} estimate a 2--10
keV luminosity of $1.7\times 10^{40}$ ergs s$^{-1}$ for X41.4+60 during the
April 2004 \emph{XMM-Newton} observation of the source. This is smaller than our
estimate of $2.5\times 10^{40}$ ergs s$^{-1}$, but this is expected given the
presence of the other ULX, which also contributes to the flux we measure from
this source.

\begin{figure}
\centering
\includegraphics[width=1.\columnwidth]{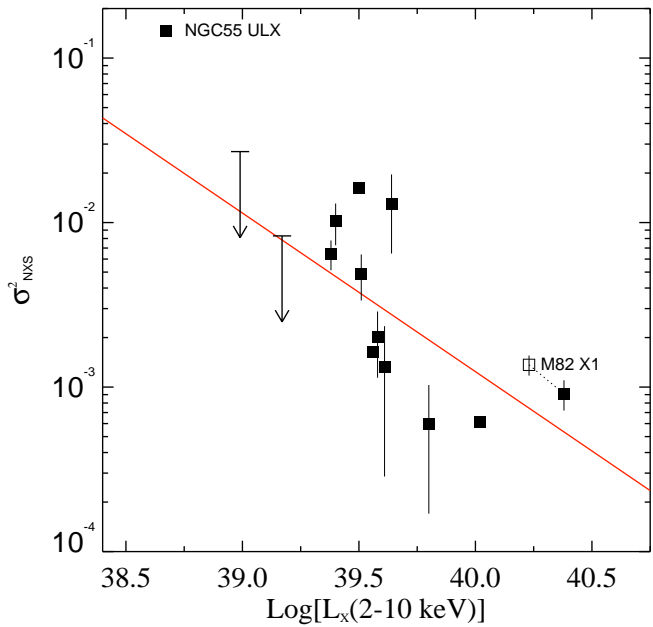}
\caption{Normalised excess variance versus log($\rm{L_{2-10~keV}}$) for the 
ULXs.  Arrows indicate the 90\% confidence upper limits on $\rm{\sigma^{2}_{NXS}}$ for
the sources with negative excess variance estimates. The point plotted with an
empty square indicates the M\,82 X--1 measurement when corrected for the contribution
of X42.3+59 to the observed count rate (see text for details). The shift of M\,82 X--1
is shown as a dotted line. The solid line
indicates the best-fit to the data  (excluding  NGC\,55 ULX).}
\label{fig:sigmalum}
\end{figure}

\subsection{The ``variability--luminosity" relation of ULXs}\label{sec:ULXrel}

Fig.~\ref{fig:sigmalum} shows $\rm{\sigma^{2}_{NXS}}$ as function of
log($\rm{L_{2-10~keV}}$) for the ULXs in the sample. The arrows indicate the
90\% confidence upper limits on the (intrinsic) excess variance of the two sources
with negative $\rm{\sigma^{2}_{NXS}}$ estimates. The X--ray variability
amplitude appears to decrease with increasing X--ray luminosity. To a large
extent, this trend is driven  by the  NGC\,55 ULX data. However, this source
shows ``dipping" episodes in its variability, which enhance its variability
amplitude \citep{Stobbart04}. Similar events have not been observed in other
ULXs. 

To investigate the significance of the apparent ``variability--luminosity"
relation in Fig.~\ref{fig:sigmalum} we fitted the [log($\rm{\sigma^{2}_{NXS}}$),
log($\rm{L_{2-10~keV}}$)] data  with a straight line of the form
log($\rm{\sigma^{2}_{NXS}}) = a + b\times$log($\rm{L_{2-10~keV}}$). Given the presence
of upper limits in two objects, we used the Buckley-James regression method  as
implemented in the software package ASURV \citep{Isobe86}. Since the NGC\,55 ULX
variability properties may be somewhat ``anomalous" among ULXs, we excluded this
source from the fit. 

The best-fit slope value is $a=-1.0\pm 0.4$ and is significantly different from
zero at the 2.5$\sigma$ level. The best-fit is indicated by the solid line in
Fig.~\ref{fig:sigmalum}. The point plotted with an empty square in
Fig.~\ref{fig:sigmalum} indicates the  M\,82 X--1 measurements, when ``corrected"
for the contribution of X42.3+59 to  the observed count rate. We have adopted
the 2--10 keV luminosity measurement of \citet{Feng07} for X41.4+60. Furthermore, we have
increased our excess variance measurement by a factor of 1.5. This is based on the fact
that the ``area A" and ``area A+B" PSDs of  \citet{Feng07} show a flat PSD at
low frequencies, whose normalisation is higher by $\sim 1.5$ in the case of
``area A" PSD. 
In the following figures we indicate only the ``corrected" data for M\,82 X--1,
albeit with a different symbol than the rest of the ULXs data.

\subsection{Comparison with AGN}

\begin{figure}[!h]
\centering
\includegraphics[width=1.\columnwidth]{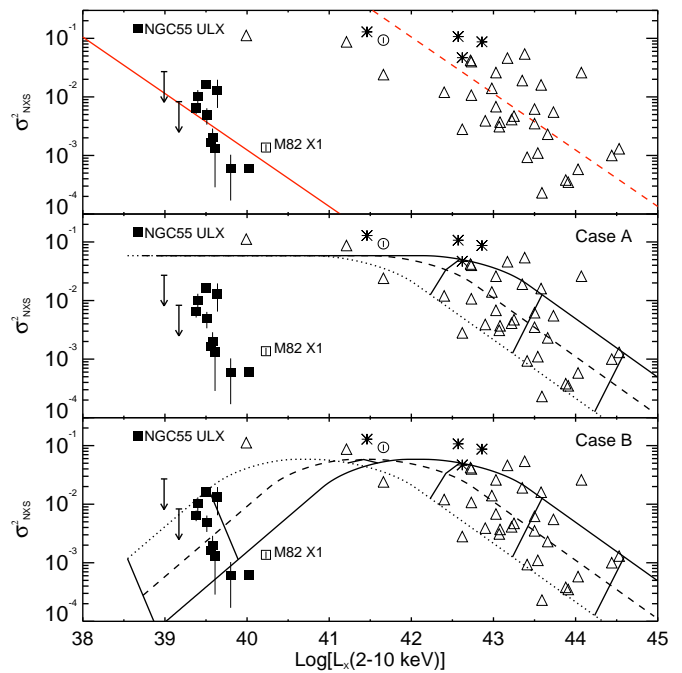}
\caption{Normalised excess variance versus log(L$_{\rm 2-10 keV}$) for ULXs
(black squares, and open square for the M\,82 X--1 data) and AGN (open
triangles). IMBHs reported by \citet{Miniutti09} are included as black stars and
POX\,52  (reported here) as an open circle. Top panel: the solid line indicates
the best fit to the ULX data, and the dashed line indicates the same line
shifted by +4 along the x--axis. Middle panel: The lines indicate the {\it case
A} model ``variability-luminosity" relation for $\dot{m}_{\rm Edd}=0.03$ (dotted
line), 0.1 (dashed line) and 0.3 (continuous line). The short  solid lines
between the model curves indicate  black hole masses of 
$5\times10^{6},~5\times10^{7}$, and $~5\times10^{8}~M_{\odot}$, from left to
right. Bottom panel: The lines indicate the {\it case B} model
``variability-luminosity" relation, for the same accretion rates. As above, the
short solid  lines within the model curves indicate black hole masses of 
$10^{3},~10^{4},~5\times10^{5},~5\times10^{6},~5\times10^{7},$ and 
$~5\times10^{8}~M_{\odot}$, from left to right.}
\label{fig:ulxagn}
\end{figure}

The top panel in Fig.~\ref{fig:ulxagn} shows the ``variability--luminosity" plot
for the ULXs (filled squares in all panels, except for the M\,82 X--1 data which
are shown with an open square;   arrows indicate 90\% confidence upper limits) and
AGN, using the data of \citet{O'Neill05} (open triangles). We also added to this
plot the [$\rm{\sigma^{2}_{NXS}}, \log({\rm L_{2-10 keV}})]$ data for the 4 AGN
with IMBHs from \citet{Miniutti09} (black asterisks). The open circles in all
panels indicate the POX\,52 data. The variability amplitude of this source  is
comparable to the amplitude of the two lowest luminosity objects in the
\citet{O'Neill05} sample and with the amplitude of the lowest luminosity object
in the \citet{Miniutti09} sample. In fact, the addition of the POX\,52 and the
four IMBHs data in the plot strengthens the possibility that the AGN
``variability--luminosity" relation may flatten at luminosities lower than $\rm{\sim
10^{42} ergs~s^{-1}}$. 

The solid line in the top panel of Fig.~\ref{fig:ulxagn}  indicates the
best-fit line for the ULXs data while the dashed line indicates the same line
shifted by $\Delta \log(L_{\rm 2-10 keV})=4$. The shifted line describes rather
well the AGN ``variability-luminosity" relation. In fact,  we used the
``ordinary least squares bisector" method of \citet{Isobe90} to fit the AGN data
(in the log--log space) with a straight line. The best-fit slope was $-1.22\pm
0.16$ which is  consistent with the best-fit slope for ULXs ($b_{\rm AGN}-b_{\rm
ULXs}= 0.2\pm 0.4$). Therefore, the variability amplitude may indeed decrease
with increasing luminosity in a similar way for AGN and ULXs. 

Despite this similarity, Fig.~\ref{fig:ulxagn} also indicates that the ULX data
are {\it not} consistent with the AGN data. For a given luminosity, even if the
AGN ``variability -- luminosity" relation flattens below $\rm{\sim
10^{42} ergs~s^{-1}}$, the ULX variability amplitude is at least 10 times lower than expected
when we extrapolate  the AGN ``variability--luminosity" relation to lower
luminosities.  Only NGC\,55 ULX appears to be consistent  with the AGN data. 

\subsection{Determination of  model ``variability--luminosity" relations}
\label{sec:models}

To compare in a quantitative way the ULXs and AGN  ``variability--luminosity"
relationship it is necessary to derive  the ``excess variance - luminosity"
relationship for AGN at low luminosities, and then  compare it  with the
observed relationship for the ULXs. In this way we will be able to investigate
if and for which physical parameters (i.e. M$_{\rm BH}$, and accretion  rate in
units of the Eddington limit, $\dot{m}_{\rm Edd}$) the ULXs data will be in
agreement with  the ``model" AGN  ``variability--luminosity" relationships.

The bolometric luminosity emitted by an AGN is $L_{\rm bol}=\dot{m}_{\rm
Edd}L_{\rm Edd}=1.3  \dot{m}_{\rm Edd}10^{38}({\rm M_{BH}/M_{\odot}~ergs~s^{-1}})$.
If $k_{bol}$ is the  X--ray to $L_{\rm bol}$ conversion factor then,

\begin{equation}
{\rm L}_{\rm 2-10 keV}=k_{bol}1.3 \dot{m}_{\rm Edd}10^{38} 
({\rm M_{BH}/M_{\odot}}) \hspace{0.2cm} {\rm ergs~s^{-1}}
\end{equation} 	

\noindent \citep [in all the calculations below, we adopted the $k_{bol}-{\rm
L_{Edd}}$ relationship given by][]{Lusso09}. On the other hand, the observed 
excess variance, estimated from a light curve of length T and bin size $\Delta
t$, is an approximate measure of the integral:

\begin{equation} 
\sigma^2_{\rm NXS}=\int_{\nu_{\rm min}}^{\nu_{\rm max}} {\rm P}(\nu)\rm d\nu ,  
\end{equation} 	

\noindent where P$(\nu)$ is the intrinsic power spectrum normalised to the
square of the light curve mean, $\nu_{\rm min}=1/T$ and   $\nu_{\rm
max}=1/(2\Delta t)$. Equation (5) results from Parseval's theorem for Fourier
series, and from the fact that the mean value of  the periodogram (i.e. the
square of the discrete Fourier transform of the light curve) is (approximately)
equal to P$(\nu)$ \citep[e.g. see discussion given in Section 2.2 of][]{Vaughan03}.  

The excess variance can be associated with the BH mass and accretion rate if one
assumes a PSD shape and certain scaling relations between the PSD characteristic
frequencies with M$_{\rm BH}$ and $\dot{m}_{\rm Edd}$. Below we present model
``variability -- luminosity" relations for AGN, assuming two rather simple
scenarios for their PSD shapes, which are based on recent power spectral studies
of AGN and GBHs.

\begin{figure}
\centering
\includegraphics[width=1.0\columnwidth]{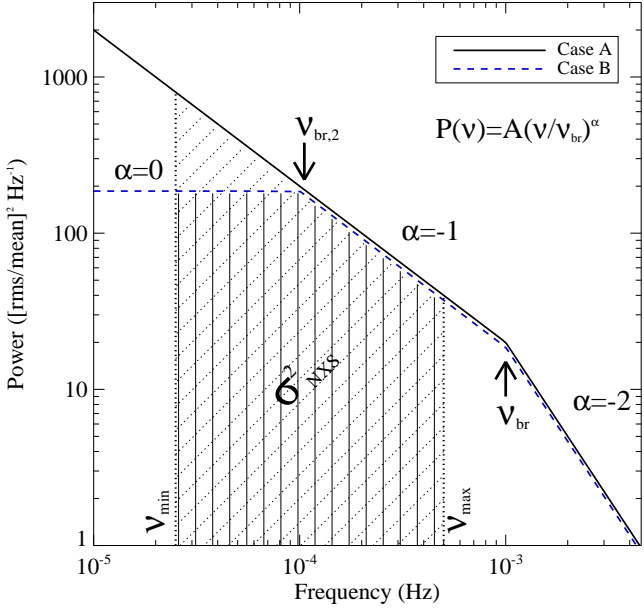}
\caption{The assumed PSD shape in {\it case A} (continuous line) and {\it case
B} (dashed line) scenarios discussed in Section~\ref{sec:models}. $\nu_{br}$ and
$\nu_{br,2}$ are the characteristic ``frequency breaks", at which the PSD slope
changes, while $\nu_{\rm min}$ and $\nu_{\rm max}$ are the minimum and maximum
frequencies sampled by the light curve segments used in this work. The
regions filled with diagonal-dotted lines (vertical-continuous lines) indicate the area below  the PSD shape
in {\it case A} ({\it case B}) scenario, which should be approximately equal to
the excess variance of the light curves. This area (and hence 
$\rm{\sigma^{2}_{NXS}}$ as well) depends on  the location of the break frequencies (determined by
black hole mass and accretion rate) with respect to $\nu_{\rm max}$ and 
$\nu_{\rm min}$ (which are fixed according to the values $T$ and $\Delta t$ of the light curves used in this work).}
\label{fig:psdmodels}
\end{figure}

{\it \underline{Case A}:} Analyses based on high quality {\it RXTE} and {\it
XMM-Newton} light curves  have shown that the AGN PSDs can be approximated by a
broken power law, with P$(\nu)=A(\nu/\nu_{\rm br})^{-1}$ and 
P$(\nu)=A(\nu/\nu_{\rm br})^{-2}$ at frequencies below and above a
characteristic frequency break,  $\nu_{\rm br}$ \citep[e.g.
][]{Markowitz03,McHardy04}. \citet{McHardy06} have demonstrated  that $\nu_{\rm
br}$ depends on both $M_{\rm BH}$ and $\dot{m}_{\rm Edd}$ as:

\begin{equation}
\nu_{\rm br}=0.003 \dot{m}_{\rm Edd}({M_{\rm BH}}/10^6~{M_{\odot}})^{-1}.
\end{equation} 	

\noindent In this case ({\it case A} model hereafter), it is straightforward to
show that:

\begin{equation}
\sigma^2_{\rm NXS}= \left\{ \begin{array}{ll}
{\rm C_{1}}\nu_{\rm br}(\nu_{\rm min}^{-1}- 
\nu_{\rm max}^{-1}), & \textrm{(if $\nu_{\rm br}<\nu_{\rm min}$)} \\
{\rm C_{1}}[{\rm ln}(\frac{\nu_{\rm br}}{\nu_{\rm min}})
-\frac{\nu_{\rm br}}{\nu_{\rm max}}+1], 
& \textrm{(if $\nu_{\rm min}<\nu_{\rm br}<\nu_{\rm max}$)} \\
{\rm C_{1}}{\rm ln}(\frac{\nu_{\rm max}}{\nu_{\rm min}}), & 
\textrm{(if $\nu_{\rm br}>\nu_{\rm max}$)}
\end{array} \right.
\end{equation} 	

\noindent where $\rm C_{1}=A\nu_{\rm br}$. Following \citet {Papadakis04} we
assumed that $C_1=0.02$. The solid line in Fig.~\ref{fig:psdmodels} indicates
the \emph{case A} PSD model, when $\nu_{\rm br}=10^{-3}$  and $A=20$
Hz$^{-1}$ (values chosen arbitrarily).  The vertical, dotted lines indicate the
$\nu_{\rm min}$ and $\nu_{\rm max}$ frequencies. The diagonal dotted lines
indicate the area below the  \emph{case A} PSD curve, and between $\nu_{\rm
min}$ and $\nu_{\rm max}$ values, which according to equation (5) should be
(approximately) equal to $\rm{\sigma^{2}_{NXS}}$. 

{\it \underline{Case B}:} A second frequency break,  $\nu_{br,2}$, below which
the PSD is roughly flat (i.e. P$(\nu)\propto \nu^{0}$), has also been observed
in at least one AGN  \citep[i.e.  Ark\,564,][]{Papadakis02,McHardy07} and in
GBHs in ``low/hard" and ``very high" states (see discussion in Section 5.1
below). The PSDs in the latter case are quite complex, usually described by a
series of Lorentzians \citep[e.g.][]{Pottschmidt03}. However,  the entire
spectral shape roughly resembles a (doubly) broken power law of the form: 
P$(\nu)=A(\nu/\nu_{\rm br})^{-2}$, for $\nu>\nu_{\rm br}$,  
P$(\nu)=A(\nu/\nu_{\rm br})^{-1}$, for $\nu_{\rm br,2}< \nu <nu_{\rm br}$ and
P$(\nu)=A(\nu_{\rm br,2}/\nu_{\rm br})^{-1}=$constant, for  $\nu<\nu_{\rm
br,2}$. In this case  ({\it case B} model hereafter), the excess variance of the
light curves should be:

\begin{equation}
\sigma^2_{\rm NXS}= \left\{ \begin{array}{ll}
{\rm C_{1}}\nu_{\rm br}(\nu_{\rm min}^{-1}- 
\nu_{\rm max}^{-1}), & \\
~~~~~~~~~~~~~~~~~~~~~~~~~~~~~~~~\textrm{(if $\nu_{\rm br,2},\nu_{\rm br}<\nu_{\rm min}$)}   & \\
{\rm C_{1}}[{\rm ln}(\frac{\nu_{\rm br}}{\nu_{\rm min}})
-\frac{\nu_{\rm br}}{\nu_{\rm max}}+1], & \\
~~~~~~\textrm{(if $\nu_{\rm min}<\nu_{\rm br}<\nu_{\rm max}$ and $\nu_{\rm br,2}<\nu_{\rm min})$} & \\
{\rm C_{1}}{\rm ln}(\frac{\nu_{\rm max}}{\nu_{\rm min}}), & \\
~~~~~~~~~~~~~~~~\textrm{(if $\nu_{\rm br}>\nu_{\rm max}$ and $\nu_{\rm br,2}<\nu_{\rm min}$)} & \\
{\rm C_{1}}[\ln(\frac{\nu_{\rm br}}{\nu_{\rm br,2}})+2-
\frac{\nu_{\rm min}}{\nu_{\rm br,2}}-\frac{\nu_{\rm br}}{\nu_{\rm max}}],  & \\
 ~~~~~~~~~~~~~~~~~~~(\textrm{if $\nu_{\rm min}<\nu_{\rm br,2}<\nu_{\rm br}<\nu_{\rm max}$}) & \\
{\rm C_{1}}[\ln(\frac{\nu_{\rm max}}{\nu_{\rm br,2}})+1-
\frac{\nu_{\rm min}}{\nu_{\rm br,2}}],  & \\
 ~~~~~~(\textrm{if $\nu_{\rm min}<\nu_{\rm br,2}<\nu_{\rm max}$ and $\nu_{\rm br}> \nu_{\rm max}$}) & \\
{\rm C_{1}}(\nu_{\rm max}-\nu_{\rm min})/\nu_{\rm br,2}, & \\ 
~~~~~~~~~~~~~~~~~~~~~~~~~~~~~~~(\textrm{if  $\nu_{\rm br,2}, \nu_{\rm br}> \nu_{\rm max}$}) & 
\end{array} \right.
\end{equation} 	

\noindent The ratio $\nu_{\rm br}/\nu_{\rm br,2}$  is usually  $\sim 10$ in
GBHs. We adopted the  \citet{Axelsson06} relation between the two break
frequencies in Cyg X--1 \footnote{Strictly speaking, equation (9) is valid for
Cyg X--1 in its ``low/hard" state. We verified that our results do not depend
strongly on the adopted relationship between $\nu_{\rm br,2}$ and $\nu_{\rm
br}$, as long as  $\nu_{\rm br,2}\sim 0.2-0.05\nu_{\rm br}$.}:

\begin{equation}
\nu_{\rm br,2} = 0.15\nu_{\rm br}^{1.2}.
\end{equation}

\noindent The dashed line in Fig.~\ref{fig:psdmodels} indicates  the \emph{case
B} PSD model, assuming a second break frequency which is 10 times smaller than
$\nu_{\rm br}$. The vertical lines between $\nu_{\rm min}$ and  $\nu_{\rm max}$
indicate the area below the the \emph{case B} PSD curve, which should be
approximately equal to $\rm{\sigma^{2}_{NXS}}$. We can now use the equations above
to construct AGN model ``variability--luminosity" relations as follows.

\subsection{The AGN model ``variability--luminosity" relations}

We considered BH mass values in the range between $10^3-10^9$ M$_{\odot}$, and
three accretion rate values, namely 0.03, 0.1 and 0.3 of the Eddington limit. 
For any given M$_{\rm BH}$ and $\dot{m}_{\rm Edd}$ values we used equation (4)
to compute the 2--10 keV luminosity of the source, and  equations (6) and (9) to
compute $\nu_{\rm br}$ and $\nu_{\rm br,2}$, respectively.  We then used
equations (7) and (8) to compute the model $\rm{\sigma^{2}_{NXS}}$ values in  {\it
case A} and  {\it case B} scenarios, respectively. The minimum and maximum
sampled frequencies in our case are $\nu_{\rm max}=1/(2\times 1000)$ Hz and
$\nu_{\rm min}=1/T_{\rm mean}$ Hz, where $T_{\rm mean}=37$ ksec, i.e. the
average length of all the segments listed in column (9) of Table ~1. The
resulting model ``variability--luminosity" relations are plotted in 
Fig.~\ref{fig:ulxagn} (middle and bottom panels).

The dotted, dashed and solid line in the middle panel of Fig.~\ref{fig:ulxagn} 
indicate the expected {\it case A} ``variability--luminosity" relations for 
$\rm{\dot{m}_{\rm Edd}=}$0.03, 0.1, and 0.3, respectively. The agreement between
these lines and the AGN data is reasonably good, indicating that both the shape
and the scatter in the  observed ``variability--luminosity" relation for AGN can
be explained if the nearby, bright Type-1 Seyferts accrete at $\sim 3-30$\% of
the Eddington limit. The flattening of the relation at low luminosities is due
to  the fact that, if equation (6) is valid, then $\nu_{\rm br}>\nu_{\rm max}$ for
objects with BH mass (luminosity) less than $\sim 2-6\times 10^5$ M$_{\odot}$
[$(0.6-4.0)\times 10^{41} erg~s^{-1}$]. Consequently,  the excess variance
should remain constant [see bottom relationship in the set of equations (7)]  for
all objects with smaller M$_{\rm BH}$ (and therefore source luminosity).

The dotted, dashed, and solid line in the bottom panel of Fig.~\ref{fig:ulxagn}
indicate the {\it case B} ``variability--luminosity" predictions (as before, we
considered the values of $\rm{\dot{m}_{\rm Edd}=}$0.03, 0.1, and 0.3, respectively).
If there is indeed a second PSD frequency  break, and equations (6) and (9) are
valid, then we expect  $\nu_{\rm br,2}>\nu_{\rm min}$ for sources  with BH mass
(luminosity) smaller than $\sim 6\times 10^5$ M$_{\odot}$ ($\rm{\sim  4\times
10^{41}ergs~s^{-1}}$). Since the  low-frequency flat part of the PSD does not
contribute to the integral defined by equation (1) as much as the $\nu^{-1}$
part does (see Fig.~\ref{fig:psdmodels}), the excess variance is expected to 
decrease with decreasing luminosity when L$_{\rm 2-10 keV}<4\times 10^{41}erg~s^{-1}$.

\begin{figure}[!t] 
\centering 
\includegraphics[width=1.\columnwidth]{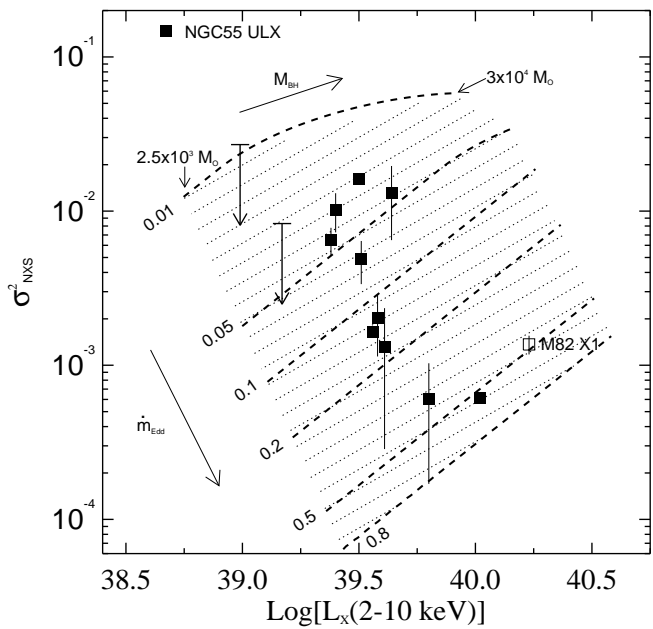}
\caption{Normalised excess variance versus log($\rm{L_{2-10~keV}}$) for the 
ULXs. The dotted-filled region indicates the area
with BH mass of $\rm{(2.5-30)\times 10^3}$  M$\rm{_{\odot}}$ and accretion rates 
of $\rm{\dot{m}_{\rm Edd}}$=(0.01-0.8).
Dashed lines indicate the  {\it case B} model for AGN-like objects with  a
BH mass of $\rm{(2.5-30)\times 10^3}$  M$\rm{_{\odot}}$ (from left to right along
each  line), and  for accretion rates of $\rm{\dot{m}_{\rm Edd}}$=[0.01, 0.05, 0.1,
0.2, 0.5, 0.8] (from top to bottom).}
\label{fig:sigmalum2} 
\end{figure}

\subsection{The comparison between ULXs and AGN revisited}\label{sec:revis}

It is clear from the middle panel of  Fig.~\ref{fig:ulxagn} that,  apart
from NGC\,55 ULX, the  ULXs data are  not consistent with the {\it case A} AGN
model predictions. On the other hand, the  bottom panel in 
Fig.~\ref{fig:ulxagn} shows that most of the ULXs data are consistent with the {\it case
B} model ``variability--luminosity" relations.

In Fig.~\ref{fig:sigmalum2} we plot again the $\rm{\sigma^{2}_{NXS}}-$log$(L_{\rm
2-10 keV})$ data for the ULXs. The dashed lines in the same plot  indicate the
{\it case B} model ``excess variance -- luminosity" relations for AGN-like
objects with  a BH mass of $\rm{(2.5-30)\times 10^3}$  M$_{\odot}$ (from
left to right along each line), and  for various accretion rates. The dashed
line on the top-left corner of the plot corresponds to $\dot{m}_{\rm Edd}=0.01$.
Moving down along the lines, the accretion rate increases 
up to  $\dot{m}_{\rm Edd}=0.8$.  The fact that most of the  ULXs data are located
within the boundaries of the diagonally dashed region of the plot  implies that 
these objects may operate like AGN, with a BH mass in the range $\rm{(2.5-30)\times
10^3}$ M$_{\odot}$,  and an accretion rate in the range 
$\rm{[0.01-0.8]}$. Regarding NGC\,2403 X--1 and NGC\,4945 X--2 (the upper limits),  their
low variability amplitude can be explained if they host a BH with a mass close
to or even lower than $2.5\times 10^3$ M$_{\odot}$.

Furthermore, the same figure can also explain  the apparent ULX
``variability-luminosity" anti-correlation. 
If the higher-luminosity objects in the sample  have systematically larger BH 
mass and accretion rate than the lower-luminosity objects, then their
variability amplitude  should also be systematically smaller, hence the ``smaller 
variability amplitude with increasing luminosity" trend we detected.

Finally, it is clear from  Fig.~\ref{fig:ulxagn} that NGC\,55 ULX is not
consistent with the {\it case B} model predictions. One could assume that this
source is more consistent with the {\it case A} model predictions, however, the
high amplitude dipping episodes seen in the light curve of this source are not
commonly seen in AGN light curves. Furthermore, if that were the case,  we would
expect that $\nu_{\rm br}>\nu_{\rm max}$ for this object, and its PSD to have a
$-1$ slope in the frequency range $\sim 10^{-4}-10^{-3}$ Hz. However, \citet{Heil09}
found an average  PSD slope of $-1.96\pm 0.04$ in this frequency
range. Therefore, the agreement of the NGC\,55 ULX data with the {\it case A}
``excess variance - luminosity"  relation must be coincidental.

\section{Discussion}

We present the results of a variability analysis of a sample of 14 bright ULXs
using 19 observations with \emph{XMM-Newton}/EPIC. We calculated their
normalised  excess variance using light-curves of 40 ksec length. Our main  aim
was to compare their ``variability--luminosity" relationship with the same
relationship for AGN. Our main results can be summarised as follows:

\begin{itemize}
\item The variability amplitude of ULXs is significantly smaller than that
expected from a simple extrapolation of the AGN ``variability--luminosity"
relationship to lower luminosities.
\item We found evidence that the variability amplitude in ULXs 
decreases with increasing 2--10 keV source luminosity. This ``variability --
luminosity" anti-correlation is similar (in slope) to what is observed in nearby
Type 1 Seyferts.
\end{itemize}

We discuss in some detail some implications of our results below.

\subsection{Are most ULXs AGN-like objects?}

The fact that ULXs show a significantly smaller variability amplitude (when
compared to the amplitude expected from an extrapolation of the AGN
``variability--luminosity" relation to low luminosities) is consistent with the 
hypothesis that ULXs are ``scaled-down" version of the nearby AGN, but only if: 
a) there two break frequencies in their PSDs, b) $\nu_{br}$  scales with BH mass
and accretion rate as in equation (6),  c) $\nu_{\rm br}/\nu_{\rm br,2} \sim
5-50$, and d) $\nu_{\rm br,2}$ (i.e. the frequency where the PSD shape changes
from a slope of 0 to $\sim -1$) is higher than $3\times 10^{-5}$ Hz. 

This band-limited noise PSD shape is commonly observed in GBHs in their
``low/hard" and ``very high" state \citep[VHS; see for example Fig.~4e and 4c 
in ][]{Klein-Wolt08}. In this case the variability is mostly limited to $\sim
1-2$ decades of temporal frequency.  In most AGN though, the PSDs resemble that
of GBHs in their ``high state": the $1/\nu$ part of the power spectrum extends
over many decades of frequency below $\nu_{\rm br}$ \citep{McHardy06}. The only
exception is  Ark\,564, where a second PSD break (to a slope flatter than $-1$)
{\it is} observed, and most of the variability is indeed limited in less than two
decades of frequency. Interestingly,  Ark\,564 is the only high accretion rate
AGN for which a good quality PSD is available at the moment. \citet{McHardy07}
suggested that this object is the AGN analogue of GBHs in VHS. Perhaps then,
just like Ark\,564, ULXs are also similar to GBHs in VHS.

Are there any indications that ULXs power spectra have  a shape consistent
with the PSD model outlined above? Recently, \citet{Heil09} presented the
results from a detailed PSD analysis of archival ULX light curves. They detected
``0 to --1" PSD breaks at frequencies higher than $3\times 10^{-5}$ Hz in
NGC\,5408 X--1 and M\,82 X--1. In NGC\,1313 X--1 they detected a ``0 to --2" break, however
the best-fit high-frequency slope has a large uncertainty ($-2.35\pm 1$). The
possibility of a --1 slope cannot be excluded, as it is just $1.35\sigma$ away
from the best fit value (note also that ``0 to --2" PSD slope breaks are
unusual in GBHs). 

On the other hand, the same authors found that the PSD of  Ho\,IX X--1, 
NGC\,1313 X--2, and NGC 55 ULX were best-fitted by a simple power-law model. The $-2$
PSD slope in NGC~55 ULX is not consistent either with the {\it case A} or {\it
case B} scenario (see also Section \ref{sec:revis}). The best-fit slope for NGC\,1313 
X--2 is also steep ($\sim -2$), but it is based on just one point \citep[see e.g.
the top left panel in Fig.~1 of][]{Heil09}, while the best-fit  slope of
Ho\,IX X--1 is flat, close to $\sim -0.5$. This may be  indicative of a
$\nu_{\rm br,2}$ break just below the lowest frequency of $10^{-4}$ Hz that \citet{Heil09}
considered. We therefore believe that the results of \citet{Heil09}  do not
disagree with the {\it case B} PSD shape we outlined in the previous section.

\subsection{Constrains on the BH mass and accretion rate of ULXs}

If indeed most of the ULXs in the sample operate like AGN, and their PSDs are
consistent with the {\it case B} scenario outlined in Section 4.5, then they 
{\it must} host a black hole with a mass of $\rm{M_{BH}\sim 2500-3\times 10^4}$
M$\rm{_{\odot}}$, which accretes at  $\sim 1-80$\% of the Eddington limit. 
Smaller BH masses with a higher accretion rate, or higher BH masses with a
smaller accretion rate cannot be consistent with most of the ULX data shown in
Fig.~\ref{fig:ulxagn}. To illustrate this point, we notice the solid line for
10$^3$M$_{\odot}$ in the bottom panel of  Fig.~\ref{fig:ulxagn}, which 
indicates the expected  $[\sigma^2_{\rm rms}, {\rm log(L_{2-10 keV}})]$
relationship in the {\it case B} model for objects with  M$_{\rm BH}=10^3$
M$\rm{_{\odot}}$  and  $\rm{0.03<\dot{m}_{\rm Edd}<0.3}$.  Clearly, this line is not
consistent with the ULX data. An increase of $k_{bol}$ by a factor of $\sim 10$
would be required to shift this line along the x-axis by the same factor, and
hence to be roughly consistent with the ULX data. This difference in $k_{bol}$ 
would imply a significant difference between the ULX and AGN spectral energy
distributions. Thus, if indeed most of the ULXs in our sample  host a BH mass
significantly smaller than $\rm{(1-10)\times 10^{3}}$ M$\rm{_{\odot}}$, then they are  not
exactly {\it like} AGN.

\citet{Wu08} also addressed the questions ``do ULXs operate
like-AGN?", and, ``if yes, what should their BH mass be"? They studied the
relationship between $\Gamma$ and X-ray luminosity in seven ULXs (four of them 
included in our sample). They found that it is similar to  what is
observed in GBHs and AGN, and that the ULX central BH mass should be
$\rm{\sim10^{4}}$ M$\rm{_{\odot}}$, in agreement with our results.  \citet{Strohmayer09}
showed that the pattern of spectral and temporal correlations in NGC\,5408
X--1 is analogous to that seen in GBHs, and argued that the BH mass range for
this system is from $\rm{(2-9)\times 10^3}$ M$\rm{_{\odot}}$. The position of the
NGC\,5408 X--1 data in Fig.~\ref{fig:sigmalum} is consistent with an object with
a BH mass of $\sim 6\times 10^3$ M$_{\odot}$  which accretes at $\sim$  4\% of
the Eddington limit, in agreement with the BH mass range of  \citet{Strohmayer09}. 
As for the M\,82 X--1 data in the same figure, they 
suggested a BH mass of $1.8\times 10^4$ M$_{\odot}$ in the system. This is
2--3 times larger than the most likely BH mass range of  ``one to a few thousand
solar masses" for this source, as estimated by \citet{Kaaret09}.
This is not a significant discrepancy, because we would need many observations  to
estimate the {\it average} variability amplitude and X--ray luminosity of the
source to compare it with the models, and therefore 
predict accurately the BH mass and accretion rate for an individual object.

\subsection{Do all ULXs in our sample operate in the ``same" way?}

Obviously, NGC\,55 ULX does not operate in the same way as the other objects in
the sample. Its large variability amplitude can be explained neither with the
{\it case A} nor the {\it case B} possibilities. This is not surprising, given
the ``dipping" episodes that have been observed in its light curve, and not in
other ULXs (or AGN). Due to the presence of these ``dips", the source light
curves have a large variability amplitude for its luminosity. 

\citet{Heil09} noticed that the strength of the intrinsic variability in
ULXs like NGC\,4559 X--1, NGC\,5204 X--1, and Ho\,II X--1 is  substantially lower
than the observed variability amplitude seen in other sources. These objects are
also included in our sample, but we found that their excess variance (i.e. their
X--ray variability amplitude) is consistent with that observed in the other
ULXs. It could be due to the technical difficulty of the estimation of the PSD
fit for faint sources like ULXs. On the other hand, we  found that 
$\rm{\sigma^{2}_{NXS}}$ is not constrained
in two other ULXs, namely NGC\,4945 X--2 and NGC\,2403 X--1.
Given the large error on the $\rm{\sigma^{2}_{NXS}}$ measurement for these two
objects, it is not clear whether their intrinsic variability amplitude is indeed
``significantly smaller" than the amplitude of the other sources. In any case
though,  if the BH mass and accretion rate of the objects in the sample are in the range
of $\rm{(2.5-30)\times 10^3}$ M$\rm{_{\odot}}$ and $\rm{\sim 1-80}$\%, respectively, we do
expect their {\it intrinsic} excess variance to differ by a factor of 
$\rm{\sim 100}$ (see  Fig.~\ref{fig:sigmalum2}). In other words, significant
differences among the  variability amplitude of ULXs do not necessarily imply
that there exist fundamental differences in their emission mechanism; they
could be due to differences in their BH mass and/or accretion rate. 

\subsection{The ``variability--luminosity" anti-correlation in ULXs}

We found evidence for an anti-correlation between the variability amplitude and
the X--ray luminosity of the ULXs we studied.  We estimated the significance of
this correlation at the 2.5$\sigma$ level. To the best of our knowledge, this
is the first time that a ``variability -- luminosity" anti-correlation has been
detected in ULXs. A similar anti-correlation was also pointed out by
\citet{Heil09} (see right panel in their Fig.~4). They parametrised the
variability amplitude by means of the power-spectrum amplitude at a given
frequency, which is significantly more difficult to measure than the excess
variance. As a result,  they could measure only upper limits on the variability
amplitude of many sources. Consequently,  when excluding the NGC\,55 ULX data
and the points with upper limits, the significance of the
``variability--luminosity" anti-correlation was substantially decreased. 

The similarity between the slope of the ``variability--luminosity"
relationship for AGN and ULXs argues in favour of the reliability of this
anti-correlation in ULXs. However, the {\it amplitude} of these two relationships is
significantly different. Under the {\it case B} scenario we have discussed
above, this anti-correlation could be explained if the high luminosity objects
in the sample have larger BH masses and higher accretion rate. In this case
they  should also have smaller variability amplitudes, as observed. 


However, there are reasons why the correlation shown in
Fig.~\ref{fig:sigmalum} may be misleading. Firstly, the luminosity of the
sources in the sample covers a rather limited range of values between
$\rm{(2.5-30)\times10^{39}erg~s^{-1}}$. In this case, even a small error in
the luminosity estimation (due to an inaccurate distance and/or flux measurement
for example) may shift the position of data points in this plot, and hence affect 
our results. We performed a numerical experiment to investigate this
effect. We used the present $(\rm{\sigma^{2}_{NXS}}-{\rm L_{2-10 keV}})$ data set
(excluding NGC\,55 ULX) to create 100 new sets. In each run, we randomly 
decreased and/or increased the luminosity of all points by a factor of 2. We
then fitted the new data set (in the log-log space) using ASURV (to take into
account the upper limits on the data points with negative excess variance
measurements, exactly as we did with the real data points)  and recorded the
best-fit slope value. In all cases, the best-ft slope was different from zero at
the (2.2-2.5)$\sigma$ level. The results from this experiment indicate that, most
probably, ``luminosity-induced" uncertainties cannot seriously affect the
observed anti-correlation.

However, the most ``serious" reason against the reliability of this
anti-correlation is the small size of our sample. A larger number of ULXs would
be necessary to confirm it. Furthermore, a conclusive test of the
``variability-amplitude" anti-correlation in ULXs will be possible when future
observations of ULXs will allow us to  determine their {\it average} excess
variance and {\it average} X--ray luminosity, and  examine if they are
anti-correlated or not.  If this correlation is confirmed it would indicate that
ULXs show a fundamental physical link between BH mass (i.e. luminosity) and
accretion rate. If this correlation is not confirmed, then what we observe in this work 
can only be a coincidence due to incompleteness of the sample.

\section{Conclusions}

The main result of this work is that the variability amplitude of ULXs is
significantly smaller than the amplitude predicted by a simple extrapolation to
low luminosities of the well established ``variability--amplitude" relationship
for the nearby bright AGN. This discrepancy can be consistent with the hypothesis
that most ULXs operate like AGN, but only if: (i) They host an IMBH of $\rm{\sim
(2.5-30)\times 10^3}$ M$_{\odot}$, (ii) their accretion rate is $1-80$\%
of the Eddington limit, and (iii) their PSDs have the band-limited noise shape
shown by GBHs in their ``low-hard" and ``very-high" state.  

We have also found evidence for an anti-correlation between the normalised excess
variance and the luminosity for ULXs. The slope is consistent with that found in
AGN but with an offset in luminosity of around four orders of magnitudes.
A larger sample of ULXs,
and the determination of their average X--ray luminosity and variability
amplitude, is necessary in order to confirm its significance.

\begin{acknowledgements}
We thank the referee for helpful comments and suggestions. We acknowledge
support by the EU FP7-REGPOT 206469 and ToK 39965 grants. This work is based on
observations with \emph{XMM-Newton}, an ESA science mission with instruments and
contributions directly funded by ESA Member States and the USA (NASA).
\end{acknowledgements}




\end{document}